\let\originalshowhyphens\showhyphens

\documentclass[sigconf]{acmart}

\copyrightyear{2026}
\acmYear{2026}
\setcopyright{cc}
\setcctype{by}
\acmConference[EASE 2026]{The 30th International Conference on Evaluation and Assessment in Software Engineering}{9–12 June, 2026}{Glasgow, Scotland, United Kingdom}
\acmBooktitle{The 30th International Conference on Evaluation and Assessment in Software Engineering (EASE '26), 9–12 June, 2026, Glasgow, Scotland, United Kingdom}
\acmPrice{}
\acmDOI{To-Be-Inserted-By-ACM}
\acmISBN{To-Be-Inserted-By-ACM}

\let\showhyphens\originalshowhyphens

\usepackage{tcolorbox}
\usepackage{multirow}
\usepackage{siunitx}
\usepackage{enumitem}
\usepackage{listings}
\usepackage{arydshln}

\usepackage{tikz}
\usetikzlibrary{shapes.geometric, arrows.meta, positioning}

\usepackage{framed}
\definecolor{shadecolor}{gray}{0.95}
\newenvironment{rqbox}{
  \par\noindent
  \setlength{\FrameRule}{0.4pt}
  \setlength{\FrameSep}{6pt}
  \begin{shaded*}\noindent
}{
  \end{shaded*}
}

\definecolor{DarkGreen}{RGB}{1,150,32}
\lstdefinestyle{PromptStyle}{
  basicstyle=\ttfamily\scriptsize,
  breaklines=true,
  frame=single,
  captionpos=b,
  rulecolor=\color{gray},
  escapeinside={(*@}{@*)},
  moredelim=**[is][\color{DarkGreen}\bfseries]{@d@}{@},
  moredelim=**[is][\color{blue}\bfseries]{@u@}{@},
  moredelim=**[is][\color{black}\bfseries]{@s@}{@}
}

\makeatletter

\newcolumntype{L}[1]{>{\raggedright\let\newline\\\arraybackslash\hspace{0pt}}m{#1}}
\newcolumntype{C}[1]{>{\centering\let\newline\\\arraybackslash\hspace{0pt}}m{#1}}
\newcolumntype{R}[1]{>{\raggedleft\let\newline\\\arraybackslash\hspace{0pt}}m{#1}}

\makeatother

\begin{document}

\title[How Compliant Are GitHub Actions Workflows? A Checklist-Based Study with LLM-Assisted Auditing]{How Compliant Are GitHub Actions Workflows?\\A Checklist-Based Study with LLM-Assisted Auditing}

\author{Edward Abrokwah}
\orcid{0009-0005-8515-8214}
\affiliation{
  \department{Department of Computer Science} 
  \institution{Trent University}
  \city{Peterborough}
  \state{Ontario}
  \country{Canada}}
\email{edwardabrokwah@trentu.ca}

\author{Taher A. Ghaleb}
\orcid{0000-0001-9336-7298}
\affiliation{
  \department{Department of Computer Science} 
  \institution{Trent University}
  \city{Peterborough}
  \state{Ontario}
  \country{Canada}}
\email{taherghaleb@trentu.ca}

\begin{abstract}
GitHub Actions (GHA) CI workflows are critical infrastructure, but current tooling offers only syntactic or heuristic checks and does not enforce documented best practices for security, maintainability, or performance. Consequently, issues like over-privileged permissions, weak secrets management, and missing failure notifications remain undetected in real-world pipelines.
This paper proposes a novel, documentation-grounded GHA compliance checklist with 30 criteria spanning four workflow sections and eight themes, and assesses Large Language Models (LLMs) for scalable compliance auditing. On 95 real-world Java workflows (2,850 assessments) using four open-weight LLMs, we find only fair agreement (Fleiss' $\kappa$ = 0.28), with systematic disagreement on structural reasoning and security-sensitive judgments. To address this, we introduce a multi-tier adjudication framework in which GPT~5 resolves model conflicts before targeted manual review, reducing verification effort by 81\% while retaining 87\% agreement with expert judgment. At scale, it reveals major compliance gaps: overall compliance is 28\%, dropping to 4\% for permission controls; Security (26\%) lags far behind Clarity (68\%). Our results show that LLMs enable scalable compliance measurement but cannot replace experts, highlighting the need for hybrid human–AI auditing and providing empirical benchmarks and guidance for defensible GHA workflow audits.
\end{abstract}



\maketitle

\section{Introduction}
Continuous Integration (CI) pipelines depend on correct, compliant workflow configurations to ensure reliable, secure, and reproducible software development. GitHub Actions (GHA), the dominant CI service on GitHub, offers powerful YAML-based automation, but this flexibility increases the risk of misconfiguration, compounded by evolving documentation and widespread third-party actions~\cite{hilton2017trade,ghaleb2025llm4ci}. Recent supply-chain attacks, such as the 2021 Codecov breach~\cite{codecov_breach}, exposed secrets from thousands of build servers, and studies show many GHA workflows are vulnerable to privilege escalation, code injection, and unauthorized token access~\cite{cycode_gha_vulns,legit_workflow_run}.

Existing validation tools check only syntactic correctness and structural well-formedness, missing workflow compliance with documented best practices~\cite{linter_semantic_gap}. Linters like \texttt{actionlint} and \texttt{yamllint} detect malformed YAML or invalid action references, but cannot assess adherence to GitHub’s security, maintainability, or performance recommendations~\cite{static_analysis_limits}. Prior work on workflow smells and configuration issues~\cite{khatami2024catching} focuses on heuristic indicators rather than normative compliance. For instance, \texttt{permissions: write-all} is syntactically valid and may not trigger smell detectors, yet it violates least privilege, leaving issues such as improper permission scoping, weak secrets management, missing failure notifications, and absent caching strategies undetected in production pipelines.
Despite GHA’s widespread adoption, non-compliance in real workflows is common, violation patterns remain unclear, and the ability of modern language models in detecting workflow compliance failures at scale is largely unknown. Addressing these gaps requires an empirical study of GHA compliance quality.

To address this gap, this paper proposes a novel, documentation-grounded checklist comprising 30 criteria across four workflow sections and eight themes, enabling automated evaluation of workflow compliance. We empirically evaluate four open-weight Large Language Models (LLMs) on 95 real-world workflows (2,850 assessments), revealing only fair inter-model agreement and systematic weaknesses in structural reasoning, security judgments, and context-dependent interpretation. Finally, we introduce a multi-tier adjudication framework combining GPT~5 dispute resolution with targeted manual review, reducing verification effort by 81\% while maintaining 87\% agreement with expert judgment. Together, these contributions provide a practical, scalable approach to measuring and mitigating compliance violations in GHA workflows.

\smallskip
\noindent\textbf{Contributions.}
This paper makes the following contributions:

\begin{enumerate}[leftmargin=2.5em]
    \item A novel, documentation-grounded checklist for GHA compliance across four workflow sections and eight themes.  
    \item Empirical evaluation of four LLMs and a hybrid {GPT~5}+human adjudication framework, revealing fair agreement, systematic weaknesses, and insights into disagreement patterns.
    \item Analysis of real-world open-source workflows, reporting compliance rates and common patterns of violation.
\end{enumerate}

\noindent\textbf{Paper Organization.}
The rest of this paper is organized as follows.
Section~\ref{sec:Background and Related Work} presents background and related work.  
Section~\ref{sec:Study Setup} presents our empirical study across four research questions (RQs).  
Section~\ref{sec:Discussion} discusses the implications of our findings.  
Section~\ref{sec:Threats to Validity} details threats to validity.  
Section~\ref{sec:Conclusion} concludes and suggests future work.

\section{Background and Related Work}
\label{sec:Background and Related Work}

\noindent\textbf{Continuous Integration (CI).}
CI automates building, testing, and deployment, enabling early detection of integration issues~\cite{Fowler_CI,valenzuela2024hidden}. Workflows are defined via YAML configuration files specifying jobs and steps, but ensuring correctness and compliance with platform best practices remains challenging.

\smallskip
\noindent\textbf{GitHub Actions (GHA).}
Since its 2019 release, GHA has seen wide adoption for its GitHub integration and extensive documentation~\cite{gha_docs} on how to set up workflow jobs with each with a clear, unique name, suitable runners, and well-defined steps using pinned third-party actions. Shell scripts (\texttt{run}) must follow safe practices; environment variables (\texttt{env}) should not hardcode secrets; inputs (\texttt{with}) should be validated; and \texttt{GITHUB\_TOKEN} permissions should be minimal. Properly structuring jobs, steps, dependencies, inputs, and permissions is essential for secure CI pipelines.

\smallskip
\noindent\textbf{Large Language Models (LLMs).}
LLMs are increasingly applied to automate compliance evaluation in GHA workflows. While they scale better than manual audits, LLMs exhibit biases, anchoring effects, and inconsistencies across samples~\cite{stureborg2024large,wang2024large,liu2024llms,zheng2023judging}, potentially overlooking novel workflow patterns. Mitigation strategies improve reliability and consistency, but understanding these limitations is crucial for safe application in CI compliance auditing~\cite{stureborg2024large,liu2024llms}.

\smallskip
\noindent\textbf{Related Studies on CI (Mis)Configurations.}
CI configurations can directly impact build time and status~\cite{ghaleb2019duration,ghaleb2022interplay,ghaleb2019noise} and are linked to mobile app success~\cite{zhou2026roleci}, underscoring the need to study how well CI workflows follow guidelines and best practices.
Prior work has examined CI configuration complexity~\cite{ghaleb2025android}, misconfigurations~\cite{gallaba2018use}, and smells~\cite{zampetti2020empirical,khatami2024catching}, as well as automated techniques to detect and repair these issues~\cite{vassallo2020configuration,zhang2022buildsonic}. Such approaches, however, largelybut these typically rely on heuristics or narrow issue taxonomies rather than systematic, documentation-grounded compliance reasoning.
Recent LLM-based methods support software quality tasks such as test smell detection~\cite{santana2025evaluating}, code smell assessment~\cite{taibi2017developers}, and repository-level audits~\cite{guo2025repoaudit}, yet their suitability for structured compliance reasoning over CI configuration guidelines remains unexplored.

\smallskip
\noindent\textbf{Related Studies on LLMs for CI.}
Recent work explores using LLMs for CI configuration, automation, and analysis. Studies show LLMs can generate GitHub Actions workflows from natural language~\cite{ghaleb2025llm4ci}, automate DevOps pipeline creation~\cite{mehta2023automated}, migrate CI across services~\cite{hossain2025cigrate}, analyze CI practices in open-source projects~\cite{chomkatek2025decoding}, and assist in failure diagnosis and remediation~\cite{xu2025logsage}. While promising, these approaches face challenges in correctness, semantic alignment, and structural reasoning, motivating our focus on normative, documentation-grounded compliance evaluation.

\smallskip
\noindent\textbf{Positioning.}  
Despite these advances, prior work has not studied CI workflow compliance auditing using structured, documentation-grounded checklists, nor analyzed inter-model agreement on compliance judgments. In the context of CI workflow compliance checking, the use of a strong adjudicating LLM to resolve model disagreements, together with targeted manual verification to establish reliable ground truth, also remains unexplored. These gaps motivate our investigation into LLM-based compliance detection and the factors driving inter-model disagreement.

\section{Empirical Analysis and Results}
\label{sec:Study Setup}
\begin{figure*}[t]
\centering
\resizebox{\textwidth}{!}{
\begin{tikzpicture}[
    box/.style={
        rectangle, rounded corners=3pt, draw,
        minimum width=2.4cm, minimum height=1.0cm,
        align=center, font=\small},
    decision/.style={
        diamond, aspect=2, inner sep=1pt,
        fill={rgb,255:red,241;green,239;blue,232},
        draw={rgb,255:red,95;green,94;blue,90},
        text={rgb,255:red,44;green,44;blue,42}, align=center},
    box-gray/.style  ={box, fill={rgb,255:red,241;green,239;blue,232},
                            draw={rgb,255:red,95;green,94;blue,90},
                            text={rgb,255:red,44;green,44;blue,42}},
    box-purple/.style={box, fill={rgb,255:red,238;green,237;blue,254},
                            draw={rgb,255:red,83;green,74;blue,183},
                            text={rgb,255:red,38;green,33;blue,92}},
    box-blue/.style  ={box, fill={rgb,255:red,230;green,241;blue,251},
                            draw={rgb,255:red,24;green,95;blue,165},
                            text={rgb,255:red,4;green,44;blue,83}},
    box-amber/.style ={box, fill={rgb,255:red,250;green,238;blue,218},
                            draw={rgb,255:red,133;green,79;blue,11},
                            text={rgb,255:red,65;green,36;blue,2}},
    box-teal/.style  ={box, fill={rgb,255:red,225;green,245;blue,238},
                            draw={rgb,255:red,15;green,110;blue,86},
                            text={rgb,255:red,4;green,52;blue,44}},
    arrow/.style={->, >=Stealth, thin,
                  color={rgb,255:red,136;green,135;blue,128}},
    lbl/.style={color={rgb,255:red,95;green,94;blue,90}, font=\scriptsize}
]

\node[box-amber] (docs)      at (0,    0)   {GHA\\Documentation};
\node[box-amber] (review)    at (3.2,  0)   {Expert Review\\(2 Co-Authors)};
\node[box-amber] (checklist) at (6.4,  0)   {30-item\\Checklist};

\node[box-gray]  (projects)  at (0,   -2.8) {8,924 Java\\Projects};
\node[box-blue]  (ghafilter) at (3.2, -2.8) {GHA Filter\\11,031 workflows\\22,990 YAML files};
\node[box-blue]  (qualfilter)at (6.4, -2.8) {Quality Filter\\$\geq$10 stars,\ $\geq$50 runs\\1,576 projects\\5,749 YAMLs};
\node[box-teal]  (sample)    at (9.9, -2.8) {Stratified Sample\\2,850 checklist items\\95\% CI,\ 10\% MoE};

\node[box-gray]  (input)     at (9.9, -1.4) {Input YAML\\Workflows};

\node[box-purple] (models) at (13.6, -1.4) {
    \begin{tabular}{@{}c@{}}
        Independent\\Model Analysis\\[2pt]\hline\\[-5pt]
        \texttt{Phi4 14B}\\[1pt]
        \texttt{Gemma-3~12B}\\[1pt]
        \texttt{mistral-0.3~7B}\\[1pt]
        \texttt{LLaMA-3.1~8B}
    \end{tabular}};
\node[decision]  (decision)  at (17.6, -1.4) {4/4 or 3/4?};
\node[box-blue]  (gpt5)      at (21.2, -1.4) {GPT~5\\Tie-Breaker};
\node[box-teal]  (output)    at (24.8, -1.4) {Final\\Compliance Labels};
\node[box-amber] (human)     at (21.2, -3.4) {Human Review};

\draw[arrow] (docs)      -- (review);
\draw[arrow] (review)    -- (checklist);

\draw[arrow] (projects)  -- (ghafilter);
\draw[arrow] (ghafilter) -- (qualfilter);
\draw[arrow] (qualfilter)-- (sample);

\draw[arrow] (checklist) -| (input);
\draw[arrow] (sample)     -- (input);

\draw[arrow] (input)    -- (models);
\draw[arrow] (models)   -- (decision);
\draw[arrow] (decision) -- node[lbl, above]{No}         (gpt5);
\draw[arrow] (gpt5)     -- node[lbl, above]{Resolved}   (output);
\draw[arrow] (gpt5)     -- node[lbl, right]{Unresolved} (human);
\draw[arrow] (human)    -| (output);

\draw[arrow]
    (decision.north)
    .. controls +(0,1.4) and +(0,1.4) ..
    node[lbl, above]{Yes}
    (output.north);

\draw[dashed, color={rgb,255:red,180;green,178;blue,170}, thin]
    (11.7, 0.75) -- (11.7, -3.85);

\node[font=\normalsize\bfseries\itshape, text={rgb,255:red,95;green,94;blue,90}]
    at (5.5,  0.85) {Checklist Derivation};
\node[font=\normalsize\bfseries\itshape, text={rgb,255:red,95;green,94;blue,90}]
    at (5.5, -3.85) {Dataset Construction};
\node[font=\normalsize\bfseries\itshape, text={rgb,255:red,95;green,94;blue,90}]
    at (18,  0.5) {Compliance Auditing Pipeline};

\end{tikzpicture}}
\vspace{-11pt}
\caption{Full study pipeline. \emph{Left}: checklist derivation from GHA documentation (top) and dataset construction from 8,924 Java projects, filtered and sampled to 2,850 checklist evaluations (bottom). \emph{Right}: LLM-based compliance auditing (unanimous or near-unanimous model verdicts are accepted; disagreements escalate to GPT~5 and, if still unresolved, to manual review).}
\vspace{-8pt}
\label{fig:full_pipeline}
\end{figure*}
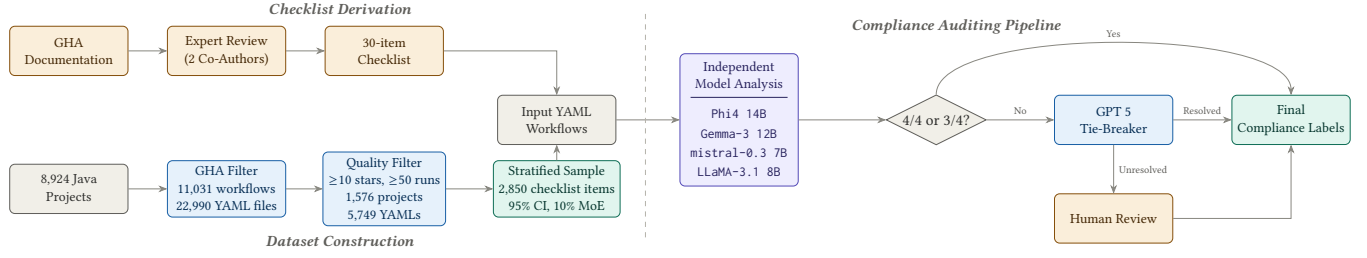

This section describes our study setup, including the construction, curation, and refinement of the proposed compliance checklist, as well as the collection and preparation of the GHA workflow dataset. We also outline the LLM-based auditing pipeline used in the study (Figure~\ref{fig:full_pipeline}), which defines how multiple models independently evaluated workflows, how disagreements were resolved, and how final compliance labels were obtained. The following subsections detail the checklist’s structure, its grounding in official GHA documentation, and the data sources, selection criteria, and preprocessing steps used to ensure consistent, reproducible, and fair evaluation across all Large Language Models.

\subsection{\textbf{RQ1. What are the core compliance criteria for GHA workflow jobs?}}
\label{RQ1_checklist}

\subsubsection{\textbf{Motivation}}
Workflow jobs are central to GHA CI pipelines, coordinating build, test, and deployment. Yet no prior work systematically defines actionable, documentation-grounded compliance criteria for jobs and steps. Existing studies highlight CI antipatterns but lack a structured checklist for reliable human or automated audits. Misconfigurations therefore persist, causing security, performance, and reliability issues, motivating our systematic approach to defining and structuring workflow job compliance criteria.

\subsubsection{\textbf{Approach}}
We derived a structured compliance checklist using a three-phase methodology: documentation review,iterative refinement and structural organization.

\begin{itemize}[leftmargin=1.55em]
    \item \textbf{\textit{Documentation Review.}} We systematically analyzed GitHub Actions’ official documentation, including workflow syntax references, guidelines, and specifications~\cite{gha_docs}. From these, we extracted required configuration elements (e.g., runner selection, permission scopes) and recommended practices (e.g., action pinning, caching strategies) for workflow jobs. Each candidate criterion was mapped to authoritative documentation to ensure platform-defined grounding. Criteria were derived only from statements with directive terms (e.g., “\textit{must}”, “\textit{should}”, “\textit{avoid}”) to capture enforceable or recommended practices. Each criterion was recorded with its source and labeled as required or recommended to support traceability and consistency.
        
    \item \textbf{\textit{Iterative Refinement.}} Two co-authors with CI expertise collaboratively refined the checklist over three iterations. In each round, we applied it to randomly sampled pilot workflows spanning single- and multi-job pipelines and varied runner configurations. Pilot workflows were selected to maximize diversity, helping uncover edge cases and ambiguities in the criteria. Items were assessed for interpretability, measurability, and cross-project applicability. Problematic items were revised or removed if they could not be clearly measured or observed, or if they duplicated higher-level criteria without adding meaningful value. Disagreements were resolved through discussion, using GitHub documentation as the final reference.
        
    \item \textbf{\textit{Structural Organization.}} The final checklist was organized along two dimensions. Each criterion was mapped to its corresponding workflow \textbf{\textit{Section}} (\textit{workflow}, \textit{jobs}, \textit{permissions}, or \textit{steps}) following the GHA schema and assigned a primary compliance \textbf{\textit{Theme}} (e.g., \textit{Security}, \textit{Error/Failure Handling}, \textit{Modularity}, etc.). This dual structure enables both targeted section-level auditing and thematic compliance analysis.
\end{itemize}

Checklist derivation followed established guidelines for operationalizing qualitative concepts in empirical software engineering~\cite{stol2016grounded}. One co-author extracted candidate criteria from the documentation, which were then reviewed and verified with the second co-author through collaborative discussion, resolving any disagreements by consulting official GHA documentation.
Each criterion was formalized as explicit binary decision rules to enable consistent \textit{yes/no} assessment. For example, “\textit{proper secrets management}” was formalized as requiring secrets to be referenced via \texttt{\$\{\{ secrets.NAME \}\}}, never hardcoded, and scoped to individual steps rather than job-level environments. Similarly, “\textit{failure handling}” was formalized as requiring explicit failure conditions (e.g., \texttt{if: failure()}) or notification steps triggered on job failure.

\vspace{3pt}
\subsubsection{\textbf{Findings}}
Table~\ref{tab:checklist} presents the proposed compliance checklist, consisting of 30 criteria organized by workflow sections and thematic concerns. The checklist was derived from official GitHub Actions documentation and defines the compliance dimensions evaluated in this study.

\begin{table*}[ht]
    \centering
    \caption{Compliance checklist for GHA workflows: 30 criteria organized by compliance sections and themes}
    \vspace{-8pt}
    \label{tab:checklist}
    \resizebox{\textwidth}{!}{
    \begin{tabular}{@{}p{2.6cm}p{3.7cm}p{.7cm}p{12cm}@{}}
    \toprule
    \textbf{Section} & \textbf{Theme} & \textbf{ID} & \textbf{Criterion} \\
    \midrule
    \multirow{3}{*}{\textbf{Workflow (W)}} 
       & Error/Failure Handling & W1 & Workflow should handle failures properly and provide notifications. \\
       & Environment            & W2 & Workflow should use documented and supported runner environments. \\
       & Security               & W3 & Workflow should follow security and maintainability best practices. \\
    \midrule
    \multirow{10}{*}{\textbf{Jobs (J)}} 
       & Clarity                & J1  & Job names should be clear and unique. \\
       & Clarity                & J2  & All jobs must be defined properly in the main jobs block. \\
       & Error/Failure Handling & J3  & Jobs should enable runner debug logging to allow better diagnosis of job execution. \\
       & Environment            & J4  & Runners must be appropriate for each job. \\
       & Modularity             & J5  & Jobs should remain modular and separated (setup, test, deploy). \\
       & Modularity             & J6  & Jobs should be isolated to avoid unintended side effects. \\
       & Performance            & J7  & Dependencies and tools should be cached effectively across jobs. \\
       & Performance            & J8  & Caching strategy should be portable across environments. \\
       & Performance            & J9  & Parallelism settings should be optimized and validated for better resource usage. \\
       & Performance            & J10  & Caching must be used to reduce build time. \\
       & Security               & J11 & Unauthorized runners must not be used. \\
    \midrule
    \multirow{16}{*}{\textbf{Steps (S)}} 
       & Modularity             & S1  & Any complex \texttt{run} commands should be split into smaller steps for clarity. \\
       & Modularity             & S2  & Build/deploy commands should be split into steps with error handling and caching. \\
       & Input Validation       & S3  & Inputs should be validated or sanitized to prevent unexpected behavior. \\
       & Input Validation       & S4  & User inputs for platform parameters should be validated. \\
       & Input Validation       & S5  & Boot JDK platform inputs should be validated. \\
       & Input Validation       & S6  & Validation checks must not be disabled without justification. \\
       & Error/Failure Handling & S7  & Steps should enable debug logging to make errors clear and easily traced. \\
       & Error/Failure Handling & S8  & Command-line tools should detect and report failures properly. \\
       & Maintainability        & S9  & Repository-specific conditions should be avoided or made configurable. \\
       & Maintainability        & S10 & Weak file-change detection (e.g., git status) should be avoided. \\
       & Maintainability        & S11 & Conditional jobs should be done using native GitHub strategies like matrix filters. \\
       & Maintainability        & S12 & Conditional expressions should be documented and maintainable. \\
       & Security               & S13 & Third-party actions must be pinned to specific commits \textit{SHA}. \\
       & Security               & S14 & Reusable or third-party actions should be kept up to date with stable versions. \\
       & Security               & S15 & Steps should include dedicated static/dynamic security analysis. \\
    \midrule
    \textbf{Permissions (P)} 
       & Security               & P1  & Secrets must be stored securely (no hardcoding). \\
    \bottomrule
    \end{tabular}
    }
    \vspace{-4pt}
\end{table*}

\smallskip\noindent\textbf{Step-level configurations drive most compliance violations in GHA workflows.} 
Our analysis produced a 30-item compliance checklist across four workflow sections and eight themes. Step-level configurations account for half of the checklist (15 items), capturing fine-grained execution logic where security vulnerabilities and operational failures commonly occur. Job-level criteria (11 items) focus on build coordination, environment configuration, and performance considerations, while workflow-level (3 items) and permissions (1 item) criteria address global properties affecting the entire pipeline. This distribution aligns with prior empirical evidence showing that CI violations disproportionately stem from low-level configuration decisions~\cite{gallaba2018use,zampetti2020empirical}. Notably, step-level criteria concentrate most Input \textit{Validation}, \textit{Maintainability}, and \textit{Security} checks, highlighting that compliance failures often stem from fine-grained execution logic rather than high-level workflow structure.

\smallskip\noindent\textbf{Security dominates compliance concerns, while performance, maintainability, and modularity are equally important.}
From a thematic perspective, \textit{Security} dominates the checklist with six criteria spanning all workflow sections, covering secrets management, action pinning, runner authorization, and vulnerability exposure. Several themes are represented by four criteria each, including \textit{Performance}, \textit{Error/Failure Handling}, \textit{Input Validation}, \textit{Maintainability}, and \textit{Modularity}, highlighting that robust CI pipelines require balanced attention across multiple quality dimensions rather than optimization along a single axis. \textit{Environment} and \textit{Clarity} form foundational themes with two criteria each, capturing runner selection and configuration decisions that affect workflow portability and correctness. Unlike prior work that infers workflow smells from historical change patterns~\cite{khatami2024catching}, our checklist is explicitly grounded in GitHub documentation, supporting normative compliance auditing.

\smallskip\noindent\textbf{Several compliance themes span multiple workflow sections, necessitating cross-layer analysis.}
The checklist reveals cross-cutting patterns: security criteria appear at workflow, job, and step levels, while Modularity and Environment span multiple configuration layers. This suggests that effective compliance tooling must pair section-specific checks with cross-section reasoning. Practically, the checklist helps DevOps teams prioritize CI improvements: security criteria form an automatable baseline, while performance and maintainability practices can be added incrementally. Over half of the criteria target step-level logic, and input validation remains a frequent, poorly automated source of errors.

\vspace{-6.5pt}
\begin{rqbox}
\textbf{RQ1 Summary:}
    A documentation-based checklist of 30 GHA compliance criteria reveals that step-level configurations are the primary source of violations, while security matters span all workflow sections. This framework supports targeted auditing and underscores the need to balance workflow complexity, maintainability, and compliance.
\end{rqbox}

\subsection{\textbf{RQ2. To what extent can LLMs detect GHA compliance in open source projects?}}
\label{RQ2}

\subsubsection{\textbf{Motivation}}
Prior research investigated the use of LLMs for CI-related tasks~\cite{ghaleb2025llm4ci}, yet their ability to assess GHA workflow compliance remains unexplored. Building on the structured checklist from RQ1, we investigate whether LLMs can reliably evaluate real-world workflows against these criteria, highlighting both their potential and limitations for automated compliance auditing.

\subsubsection{\textbf{Approach}}
Our study evaluates the ability of LLMs to detect GHA workflow compliance in open-source projects using the checklist developed in RQ1. We follow the following process.

\smallskip
\noindent\textbf{Dataset Preparation.}  
We used a dataset of 8,924 Java projects spanning multiple CI services from a recent study~\cite{chopra2025multici}. Projects were filtered to include only those using GHA, yielding 11,031 workflows across 22,990 YAML files. Following prior established criteria~\cite{beller2017travistorrent}, we retained projects with at least 10 stars and 50 workflow runs to focus on non-trivial, actively used repositories. This resulted in 1,576 projects with 5,749 distinct YAML workflows.

\smallskip
\noindent\textbf{Sampling Workflows.}
To balance coverage and manual feasibility, we randomly sampled 95 workflows (95\% confidence level, $\pm 10\%$ error margin) from the filtered dataset. This sample represents $95\times30 = 2,850$ possible checklist checks per LLM.

\smallskip
\noindent\textbf{LLM Selection and Configuration.}  
We evaluated four open-weight LLMs representing diverse architectures and reasoning capabilities: 
LLaMA-3.1~8B~\cite{touvron2023llama}, 
Gemma-3~12B~\cite{team2025gemma}, 
mistral-0.3~7B~\cite{mistralai2024mistral7b}, and 
Phi-4~14B~\cite{abdin2024phi}. 
GPT~5 (a proprietary reasoning model)~\cite{openai_gpt5_2025} was additionally used as an adjudicator for disagreements. 
Open-weight models were run via Ollama~\cite{marcondes2025using} with $\texttt{temperature}=0$ to ensure deterministic and reproducible outputs and reduce sampling variance, aligning with best practices for stable LLM benchmarking~\cite{blackwell2024towards}.
In contrast, GPT~5 operates at a fixed $\texttt{temperature}=1$ (not configurable via the API), which supports controlled sampling for more robust reasoning and helps avoid degenerate outputs~\cite{pipis2025wait,openai_reasoning}.
Table~\ref{tab:llms} gives more details about the LLMs used in our study.

\begin{table}[ht]
    \centering
    \renewcommand{\arraystretch}{1.1}
    \caption{LLMs used in our study}
    \vspace{-8pt}
    \resizebox{\linewidth}{!}{
    \begin{tabular}{lp{6cm}}
    \hline
    \textbf{Model} & \textbf{Description} \\
    \hline
    LLaMA-3.1~8B~\cite{touvron2023llama} & Decoder-only model from Meta with Grouped Query Attention for long-context reasoning, fine-tuned with supervised and RL methods for helpfulness, coherence, and safety. \\
    \hline
    Gemma-3~12B~\cite{team2025gemma} & Multimodal (text+image) decoder-only transformer from Google DeepMind with 400M vision encoder, supporting 140+ languages; optimized for reasoning, summarization, QA, and vision-language tasks. \\
    \hline
    mistral-0.3~7B~\cite{mistralai2024mistral7b} & Optimized transformer with grouped query and sliding window attention, efficient for long sequences, strong on reasoning, math, and code tasks. \\
    \hline
    Phi-4~14B~\cite{abdin2024phi} & Transformer-based model trained on synthetic data, cleaned web content, academic papers, and QA datasets. \\
    \hline
    GPT~5~\cite{openai_gpt5_2025} & OpenAI’s advanced multimodal model with autonomous task execution, routing queries between a fast general engine and deep reasoning core. \\
    \hline
    \end{tabular}
    }
    \label{tab:llms}
    \vspace{1pt}
\end{table}

\smallskip
\noindent\textbf{Checklist Question Formulation and Prompting.}  
Each checklist item was converted into a structured question answerable with ``\textit{YES},'' ``\textit{NO},'' or ``\textit{NOT APPLICABLE}''. We then designed a zero-shot prompt with two roles, to ensure deterministic, section-aligned responses suitable for automated analysis (Listing~\ref{lst:prompt_template}), as follows:

\begin{itemize}[leftmargin=2.5em]
    \item \textit{\textbf{System role:}} Instructed the LLM to act as a senior DevOps expert auditing GHA workflows against the checklist.
    \item \textit{\textbf{User role:}} Provided the workflow YAML and requested structured JSON output aligned with checklist sections.
\end{itemize}

\smallskip
\noindent\textbf{LLM Evaluation and Agreement Analysis.}
Each sampled workflow was evaluated by all four LLMs, producing 11,400 total outputs. To understand reliability, outputs were categorized into three agreement bands (agreement distribution is summarized in Table~\ref{tab:agreement_distribution}):

\begin{itemize}[leftmargin=2em]
    \item \textbf{Unanimous (4/4)}: All models agree.
    \item \textbf{Near-unanimous (3/4)}: Single model disagrees; treated as strong agreement.
    \item \textbf{Split (2/2 or 2/1/1)}: Significant disagreement, indicating ambiguous or context-dependent criteria.
\end{itemize}

\begin{table}[ht]
    \centering
    \vspace{-3pt}
    \renewcommand{\arraystretch}{1.05}
    \caption{Agreement Categories Across Checklist Questions}
    \vspace{-8pt}
    \label{tab:agreement_distribution}
    \resizebox{\linewidth}{!}{
    \begin{tabular}{p{5cm}rr}
    \toprule
    \textbf{Category} & \textbf{Count} & \textbf{Percentage} \\
    \midrule
    Unanimous (4/4)        & 758   & 27\% \\
    Near-unanimous (3/4)   & 1,104 & 39\% \\
    Split (2/2 or 2/1/1)   & 988   & 35\% \\
    \bottomrule
    Total                  & 2,850 & 100\%\\
    \bottomrule
    \end{tabular}
    }
    \vspace{-3pt}
\end{table}

\begin{figure}
\begin{lstlisting}[
  style=PromptStyle,
  label={lst:prompt_template},
  caption={Prompt Template used for LLMs}
]
@u@System:@
@s@     You are a senior DevOps expert specialized in secure, efficient, and standardized CI pipeline practices. Your task is to rigorously audit GHA YAML workflow files for compliance with industry-recognized best practices. You understand both functional correctness and structural quality of CI workflows. Use a sectioned JSON format matching the checklist structure.@

@u@User:@
@s@      Audit the following GHA YAML workflow using the compliance checklist provided. Return your results as a structured JSON matching the section/question layout from the checklist. Answer each question with "YES", "NO", or "NOT APPLICABLE". Return only valid JSON, with no commentary or markdown.
@
@d@     {YAML}@
\end{lstlisting}
\vspace{-14pt}
\end{figure}

\smallskip
\noindent\textbf{Hybrid Adjudication for Split Cases.}  
When LLMs disagreed on a workflow, we used GPT~5 to parse the raw YAML files and generate a consistent input, enabling systematic resolution of split decisions. Cases that remained unresolved after GPT~5 adjudication were escalated to manual review. From these, a stratified subset of 79 cases was manually adjudicated, ensuring proportional representation across checklist items and achieving a 95\% confidence level with a $\pm 10\%$ margin of error. The remaining 345 cases were not considered for manual review and thus excluded from our subsequent analyses. This process yielded 295 individual checklist item evaluations used as ground truth for model performance assessment.

\vspace{1pt}
\subsubsection{\textbf{Findings}}
Table~\ref{tab:model_performance} presents the performance of the evaluated LLMs in terms of overall agreement rates, pairwise Cohen's $\kappa$ values, and McNemar test results, highlighting both the extent of concordance between models and statistically significant differences in their compliance judgments.

\smallskip\noindent\textbf{\emph{LLM performance varies widely.}}
{\sc Gemma-3~12B} achieved the highest agreement rate at 90\%, followed by {\sc Phi-4~14B} (71\%),{\sc mistral-0.3~7B} (70\%), and {\sc LLaMA-3.1~8B} (61\%) (Table~\ref{tab:model_performance}). This suggests that parameter count alone does not predict performance; instead, architecture
and training data appear to be stronger determinants of compliance detection capability. The 29 percentage point gap between the best and worst model further indicates that results are highly model-dependent, making single-model evaluations unreliable.

\begin{table}[ht]
    \centering
    \caption{Model Agreement Rates, Pairwise Cohen's $\kappa$, and McNemar Test Results}
    \label{tab:model_performance}
    \vspace{-8pt}
    \resizebox{\linewidth}{!}{
    \begin{tabular}{l c l c c}
    \toprule
    \textbf{Model} & \textbf{Agreement} & \textbf{Pairwise} & \textbf{Cohen's} & \textbf{McNemar} \\
                   & \textbf{Rate}      & \textbf{Model}    & \textbf{$\kappa$} & \textbf{p-value} \\
    \midrule
    
    \multirow{3}{*}{{\sc Gemma-3~12B}}
        & \multirow{3}{*}{90\%}
        & {\sc LLaMA-3.1~8B} & 0.57 & 0.02 \\
     & & {\sc mistral-0.3~7B}  & 0.57 & 0.02 \\
     & & {\sc Phi-4~14B}    & 0.70 & 0.04 \\
    \midrule
    
    \multirow{2}{*}{{\sc LLaMA-3.1~8B}}
        & \multirow{2}{*}{61\%}
        & {\sc mistral-0.3~7B}  & 0.47 & 1.00 \\
     & & {\sc Phi-4~14B}    & 0.73 & 0.73 \\
    \midrule
    
    \multirow{1}{*}{{\sc mistral-0.3~7B}}
        & \multirow{1}{*}{70\%}
        & {\sc Phi-4~14B}    & 0.40 & 0.81 \\
    \midrule
    
    {\sc Phi-4~14B} & 71\% & -- & -- & -- \\
    \midrule
    
    \multicolumn{5}{c}{\textbf{Fleiss' $\kappa$ = 0.28}} \\
    \bottomrule
    \end{tabular}
    }
    \vspace{-7pt}
\end{table}

\smallskip\noindent\textbf{\emph{Consensus across checklist questions is limited.}}  
Across the checklist, unanimous agreement occurred on 27\% of questions and near-unanimous on 39\%, leaving 35\% of cases split across models (Table~\ref{tab:agreement_distribution}). This shows that a significant fraction of checklist items are ambiguous or context-dependent, suggesting that automated auditing cannot treat model outputs as definitive and requires mechanisms to resolve unresolved items.  

\smallskip\noindent\textbf{\emph{Pairwise agreement reveals systematic divergences.}}  
Pairwise $\kappa$ and McNemar tests show uneven alignment: the highest agreement is between {\sc LLaMA-3.1~8B} and {\sc Phi-4~14B} (73\%), and the lowest between {\sc mistral-0.3~7B} and {\sc Phi-4~14B} (40\%). {\sc Gemma-3~12B} differs significantly from all models ($p<0.05$), indicating distinct heuristics for interpreting checklist items and motivating multi-LLM setups with tie-breaking or human review is necessary to ensure reliable compliance assessments. In contrast, \textsc{LLaMA-3.1~8B}, \textsc{mistral-0.3~7B}, and \textsc{Phi-4~14B} show no statistically significant pairwise differences (p > 0.05), suggesting their disagreements may stem from random variation rather than systematically distinct reasoning strategies.

\begin{table*}[ht]
    \centering
    \renewcommand{\arraystretch}{1.04}
    \caption{Model Performance on Manual Validation (n=295). Bold represents the results of the highest performing model.}
    \vspace{-8pt}
    \label{tab:model_performance_on_manual_validation}
    \resizebox{\textwidth}{!}{
    \begin{tabular}{l S S S S S S S S S S S S S S}
    \toprule
    \textbf{Model} & \textbf{Acc} & \textbf{Pr.$_{macro}$} & \textbf{Re$_{macro}$} & \textbf{F1$_{macro}$} 
    & \textbf{Pr.$_Yes$} & \textbf{Re$_Yes$} & \textbf{F1$_Yes$} 
    & \textbf{Pr.$_No$} & \textbf{Re$_No$} & \textbf{F1$_No$} 
    & \textbf{Pr.$_{NA}$} & \textbf{Re$_{NA}$} & \textbf{F1$_{NA}$} \\
    \midrule
    Manual & 1.000 & 1.000 & 1.000 & 1.000 
    & 1.000 & 1.000 & 1.000 
    & 1.000 & 1.000 & 1.000 
    & 1.000 & 1.000 & 1.000 \\

    \noalign{\vskip 1pt}
    \hdashline
    \noalign{\vskip 2pt}

    GPT5 & 0.475 & 0.405 & 0.383 & 0.387 
    & 0.167 & 0.264 & 0.204 
    & \bfseries 0.410 & 0.302 & \bfseries 0.348 
    & 0.640 & \bfseries 0.582 & 0.609 \\
    
    mistral-0.3~7B & 0.356 & 0.303 & 0.289 & 0.280 
    & 0.136 & 0.302 & 0.187 
    & 0.184 & 0.132 & 0.154 
    & 0.590 & 0.434 & 0.500 \\
    
    Phi-4~14B & \bfseries 0.485 & \bfseries 0.438 & \bfseries 0.451 & \bfseries 0.421 
    & 0.240 & 0.340 & 0.281 
    & 0.268 & \bfseries 0.491 & 0.347 
    & \bfseries 0.805 & 0.524 & \bfseries 0.635 \\
    
    LLaMA-3.1~8B & 0.220 & 0.304 & 0.314 & 0.224 
    & 0.159 & \bfseries 0.453 & 0.235 
    & 0.187 & 0.377 & 0.250 
    & 0.568 & 0.111 & 0.186 \\
    
    Gemma-3~12B & 0.254 & 0.364 & 0.318 & 0.268 
    & \bfseries 0.317 & 0.358 & \bfseries 0.336
    & 0.120 & 0.415 & 0.186 
    & 0.654 & 0.180 & 0.282 \\
    \bottomrule
    \end{tabular}
    }
    \vspace{-7pt}
\end{table*}

\smallskip\noindent\textbf{\emph{Models performance in the manually validated set.}} 
The results in Table~\ref{tab:model_performance_on_manual_validation} are computed over \textbf{295} manually validated instances, which form the ground-truth evaluation set. Phi-4 shows the best overall performance, with the highest accuracy (\num{0.485}), macro precision (\num{0.438}), macro recall (\num{0.451}), and macro F1 (\num{0.421}), indicating the most balanced performance across classes. GPT5 ranks second with a macro F1 of \num{0.387}, limited by weak performance on the \textit{YES} class (Precision = \num{0.167}, F1 = \num{0.204}), despite strong performance on the \textit{N/A} class (F1 = \num{0.609}).
Across all models, the \textit{YES} class has the lowest F1 scores, peaking at \num{0.336} (Gemma-3), with precision consistently low ($\leq$ \num{0.317}). For the \textit{NO} class, Phi-4 reaches the highest recall (\num{0.491}) with an F1 of \num{0.347}, while other models perform worse and less consistently. The \textit{N/A} class is the most reliably predicted, with all models achieving their highest precision on this class, especially Phi-4 (Precision = \num{0.805}, F1 = \num{0.635}), though recall varies widely (e.g., \num{0.111} for LLaMA-3.1). Overall, models perform best on \textit{N/A} and consistently struggle to detect \textit{YES} instances in the manually annotated evaluation set.
Section- and theme-level agreement rates reported below are computed from the same 2,850 assessments and are available in full in the replication package~\cite{our_replication_package}.

\smallskip\noindent\textbf{\emph{Section-level reliability is uneven.}}  
Agreement is highest for Workflow items (74\%), followed by Steps (68\%), Jobs (62\%), and lowest for Permissions (33\%). LLMs handle global workflow rules more consistently than fine-grained logic, highlighting that auditing pipelines should incorporate human or advanced-model review for critical or security-sensitive sections, particularly permissions.  

\smallskip\noindent\textbf{\emph{Theme-level trends highlight strengths and weaknesses.}}  
Models achieved the strongest agreement on Modularity (91\%) and Maintainability (73\%), moderate agreement on Performance (68\%) and Error/Failure Handling (64\%), and weakest agreement on Environment (41\%) and Security (55\%). This pattern suggests that structural, deterministic rules are easier for LLMs to assess, while context-sensitive or configuration-dependent criteria require careful human-in-the-loop verification in practice.  

\smallskip\noindent\textbf{\emph{Overall, LLMs alone are insufficient.}}  
Although some models perform well individually, their collective outputs show only fair agreement (Fleiss’ $\kappa$ = 0.28) with many split cases. LLMs can detect straightforward compliance issues but cannot reliably judge ambiguous or context-dependent checklist items. This highlights the need for hybrid pipelines that combine multiple LLMs, tie-breaking via {\sc GPT~5}, and targeted human review to ensure trustworthy, reproducible compliance auditing in real-world DevOps settings.

\vspace{-12.5pt}
\begin{rqbox}
\textbf{RQ2 Summary:}
    LLM performance on GHA compliance auditing is inconsistent, with {\sc Gemma-3~12B} reaching 90\% agreement but 35\% of criteria still requiring adjudication, showing that human oversight remains essential for reliable auditing.
\end{rqbox}

\subsection{\textbf{RQ3. Why do LLMs struggle to agree in GHA compliance auditing?}}
\label{RQ3}

\subsubsection{\textbf{Motivation}}
RQ2 revealed that LLMs evaluate GHA compliance inconsistently, with disagreements that are systematic rather than random—some models over-flag certain rule categories while under-detecting others. This highlights a key challenge: resolving conflicting LLM judgments in a principled way. Existing AI-assisted auditing approaches rarely address multi-model disagreement, often assuming a single authoritative model or using majority voting, which is inadequate when models differ due to heuristics, reasoning styles, or interpretations of compliance criteria.

\subsubsection{\textbf{Approach}}
RQ3 builds on the multi-tier adjudication strategy from RQ2 to examine why LLMs struggle with GHA compliance. Our approach combines automated adjudication with stratified manual validation to resolve disagreements among weaker models.

\smallskip
\noindent\textbf{Automated adjudication with \textsc{GPT~5}.} 
We employ \textsc{GPT~5} as a dispute resolution tool for three key reasons:
\begin{enumerate}[leftmargin=2.5em]
    \item Its larger training corpus and parameter count provide superior reasoning compared to open-weight models~\cite{gupta2026reliabilitybench}.
    \item It reduces manual review burden from 424 to 79 items (81\% reduction), making large-scale validation feasible.
    \item Stratified manual verification of 79 items shows 87\% agreement with expert judgment, confirming its effectiveness.
\end{enumerate}
Importantly, \textsc{GPT~5} was \emph{not} treated as ground truth; instead, it amplified consensus among weaker models, with manual review providing independent validation. This mirrors judicial review, where higher courts resolve lower-court disagreements, subject to final expert adjudication~\cite{amiri2025enhancing}.

\smallskip
\noindent\textbf{Stratified manual validation.} 
To assess GPT~5 adjudication reliability, we manually review a statistically representative subset of 79 items (19\% of flagged cases), following empirical LLM evaluation guidelines~\cite{wagner2025towards}. The sample guarantees coverage of:
\begin{itemize}[leftmargin=2em]
    \item \textbf{\textit{All 30 checklist criteria:}} At least two disagreements per criterion.
    \item \textbf{\textit{All disagreement patterns:}} Single-model disagreements, systematic multi-model errors, and borderline context-dependent cases.
    \item \textbf{\textit{All compliance categories:}} Security, Clarity, Performance, Modularity, Environment, Input Validation, Maintainability, and Error Handling, proportionally represented.
\end{itemize}
Each sampled item is manually reviewed by a blinded author using the original checklist rules. Disagreements with GPT~5 are resolved through discussion with a second author and reference to official GHA documentation. This ensures that systematic errors in GPT~5 adjudication would be detectable.

\smallskip
\noindent\textbf{Validation outcomes.} 
The 79-item stratified sample confirms that GPT~5 adjudication aligns closely with expert judgment (87\% agreement) across all criteria types. The absence of systematic misclassification patterns supports GPT~5’s reliability as a first-tier dispute resolver, while final adjudication of any remaining ambiguous cases ensures high-quality, trustworthy compliance judgments.

\vspace{2pt}
\subsubsection{\textbf{Findings}}
Table~\ref{tab:LLM_Why_Disagree} outlines the key reasons and underlying causes of disagreement among LLMs in GHA compliance auditing, along with corresponding findings from manual validation and the relative frequency of each issue.

\begin{table*}[ht]
    \centering
    \caption{Summary of reasons for LLM disagreement in GHA compliance auditing}
    \vspace{-8pt}
    \label{tab:LLM_Why_Disagree}
    \resizebox{1\textwidth}{!}{
    \begin{tabular}{@{}p{3cm}p{4cm}p{5cm}p{5cm}p{2.2cm}@{}}
    \toprule
    \textbf{Theme} & \textbf{Main Issue} & \textbf{Why LLMs Disagreed} & \textbf{Manual Check Findings} & \textbf{Share of Disagreements (\%)} \\
    \midrule
    
    \textbf{Job Structure \& Runners} &
    Missing jobs, unclear naming, runner configuration &
    LLMs often misinterpreted reusable workflows, treated missing runners as implied non-compliance, or inconsistently judged job completeness. &
    Manual review revealed some workflows lacked full job configuration, others used valid runners (e.g., \texttt{ubuntu-latest}), and a small fraction used unauthorized runners. &
    \multicolumn{1}{r}{\textbf{23.1}} \\
    \midrule
    
    \textbf{Inputs \& Conditional Logic} &
    Input validation and conditional expressions &
    LLMs treated the absence of inputs or conditions as violations, assuming validation logic was required. &
    Many workflows legitimately defined no inputs or conditions, making validation not applicable. &
    \multicolumn{1}{r}{\textbf{19.2}} \\
    \midrule
    
    \textbf{Security Practices} &
    Secrets handling and security scanning &
    LLMs over-flagged missing secrets or scanning, assuming all workflows must include security checks regardless of context. &
    No secrets were hardcoded; however, many workflows omitted security scanning entirely. &
    \multicolumn{1}{r}{\textbf{13.5}} \\
    \midrule
    
    \textbf{Error/Failure Handling} &
    Detecting failures or explicit error handling &
    LLMs assumed explicit failure-handling blocks were required, overlooking tools that fail on non-zero exit codes. &
    Gradle correctly failed on errors, but several workflows lacked explicit logging or failure notifications. &
    \multicolumn{1}{r}{11.5} \\
    \midrule
    
    \textbf{Caching \& Performance Optimization} &
    Use or absence of caching mechanisms &
    LLMs disagreed on when caching was required or misidentified existing caching behavior. &
    Some workflows used effective caching, while others had incomplete or missing caching. &
    \multicolumn{1}{r}{11.5} \\
    \midrule
    
    \textbf{Action Pinning \& Reusable Workflows} &
    Pinning third-party or local actions &
    LLMs incorrectly required SHA pinning for local reusable workflows and evaluated tags and SHAs inconsistently. &
    Local reusable workflows cannot be SHA-pinned by design; third-party actions varied in correctness. &
    \multicolumn{1}{r}{7.7} \\
    \midrule
    
    \textbf{Change Detection} &
    Brittle vs. dynamic file change detection &
    LLMs sometimes flagged detection logic as brittle or assumed alternatives were required. &
    Manual review confirmed mixed practices, including both robust and brittle approaches. &
    \multicolumn{1}{r}{5.8} \\
    \midrule
    
    \textbf{Modularity \& Maintainability} &
    Step structure and decomposition &
    LLMs occasionally flagged simple commands as insufficiently modular. &
    Simplicity did not reduce maintainability in these cases. &
    \multicolumn{1}{r}{3.8} \\

    \bottomrule
    \end{tabular}}
    \vspace{-1pt}
\end{table*}

\smallskip
\noindent\textbf{Automated adjudication resolves most, but not all, model disagreement.}
Of the 988 checklist items where the four open-weight LLMs disagreed, GPT~5 aligned with the majority judgment in 564 cases (57\%), resolving them without human intervention. This increased the proportion of items with stable, non-contradictory labels from 65\% to 85\%. The remaining 424 cases (43\%) still required manual verification, indicating that adjudication reduces but does not eliminate ambiguity in LLM-based compliance auditing.

\smallskip
\noindent\textbf{Disagreement concentrates on a small number of recurring compliance themes.}
Manual analysis of a stratified sample of 79 unresolved cases shows that disagreement is not uniformly distributed across the checklist. The most frequent source of conflict concerns \textit{Job Structure \& Runners} (23.1\%), where models inconsistently interpreted missing jobs, reusable workflows, and runner specifications. A second major source is \textit{Inputs \& Conditional Logic} (19.2\%), where LLMs often assumed validation or conditional checks were required even when workflows exposed no inputs or conditions, leading to systematic over-flagging.

\smallskip
\noindent\textbf{Security-related criteria trigger systematic but inconsistent risk judgments.}
Security practices account for a substantial portion of disagreement, spanning \textit{Secrets and Scanning} (13.5\%) and \textit{Action Pinning \& Reusable Workflows} (7.7\%). LLMs frequently over-reported violations by assuming that all production workflows must include security scanning or SHA-pinned actions, even when such requirements were context-dependent or inapplicable (e.g., local reusable workflows). These conflicts reflect differences in implicit risk tolerance rather than random error.

\smallskip
\noindent\textbf{Operational conventions expose limits of context-free reasoning.}
Criteria related to \textit{Error/Failure Handling} (11.5\%) and \textit{Caching \& Performance Optimization} (11.5\%) reveal a recurring issue: LLMs often expected explicit error-handling logic or caching steps, overlooking implicit guarantees provided by tools like Gradle or existing cache setups. Similarly, \textit{Change Detection} (5.8\%) disagreements occurred when models inferred brittleness without sufficient contextual evidence. These cases require understanding tool semantics and workflow intent, which is hard to infer from YAML alone.

\smallskip
\noindent\textbf{Structural clarity rarely causes disagreement.}
Only a small fraction of conflicts involved \textit{Modularity \& Maintainability} (3.8\%), where LLMs occasionally flagged simple or compact step definitions as insufficiently modular. Manual review confirmed that, in most such cases, simplicity did not reduce clarity or maintainability, explaining the low frequency of disagreement for these criteria.

\smallskip
\noindent\textbf{Overall, disagreement is widespread but highly structured.}
Across the checklist, 27 of 30 compliance criteria (90\%) had at least one case where none of the five LLMs reached consensus. These conflicts are dominated by structural interpretation, conditional applicability, and security judgment, rather than uniformly affecting all checklist items. This indicates that LLM disagreement in GHA compliance auditing is partially systematic and predictable, highlighting specific areas where automated auditing needs adjudication or human oversight.

\vspace{-10pt}
\begin{rqbox}
\textbf{RQ3 Summary:}
    LLM disagreement in GHA compliance auditing is partially systematic: \textsc{Gemma-3~12B} shows distinct, statistically significant divergence from all other models, while disagreements among the remaining three models appear less consistent, driven by structural reasoning, context dependence, and security judgment.
\end{rqbox}

\subsection{\textbf{RQ4. To what extent do open-source GitHub workflows adhere to GHA compliance checklist?}}
\label{RQ4}

\subsubsection{\textbf{Motivation}}
Our checklist captures GitHub’s recommendations for secure, maintainable, and reliable CI workflows, but it does not show whether open-source projects actually follow them. GHA’s flexibility lets maintainers favor convenience, historical configurations, or project-specific needs over formal compliance, leaving real-world adherence uncertain. Measuring actual compliance establishes a baseline for CI quality, distinguishes widely adopted from routinely ignored practices, and reveals systematic gaps that may need better tooling, documentation, or community education. This RQ therefore investigates the extent to which open-source GHA workflows align with our validated checklist.

\subsubsection{\textbf{Approach}}
To systematically evaluate GHA workflow compliance, we employed a multi-model consensus framework that balances automated assessment with targeted manual validation. For each checklist question derived from RQ2, every model produced one of three judgments: \textit{YES}, \textit{NO}, or \textit{NOT APPLICABLE (N/A)}. The interpretation of these judgments depends on the checklist item's intent: for security-positive checks (e.g., ``Is the action pinned to a commit SHA?''), \textit{YES} indicates compliance; for security-negative checks (e.g., ``Are secrets exposed in plaintext?''), \textit{NO} indicates compliance.
Items flagged as \textbf{N/A} were excluded from compliance frequency calculations, as they do not apply to the workflow in question.
Overall, 43\% of checklist evaluations were marked N/A, with the highest rates on criteria S2, J8, and S9, reflecting workflows where modularity, performance optimisation, and maintainability practices were absent and those criteria could not be evaluated. The high \textit{N/A} rate indicates that many checklist criteria cover features that are not present in all workflows. This reflects the varying scope of the studied Java projects rather than non-compliance.

Given that individual model outputs can contain errors or inconsistencies, we implemented a multi-level agreement framework to enhance reliability. A workflow checklist item is classified as \textbf{\textit{compliant}} when at least one of the following conditions is met:

\begin{enumerate}[leftmargin=2.5em]
    \item Three or all four models produce the same judgment.
    \item At least two models agree with \textsc{GPT~5}, used as a reference model for its superior reasoning capabilities.
    \item Two models agree, and their shared judgment is confirmed by manual expert review.
\end{enumerate}

Items failing to meet any of these criteria are classified as \textbf{\textit{noncompliant}}. This design prioritizes automated consensus while retaining human verification for ambiguous or borderline cases, ensuring both accuracy and efficiency.
Using the validated compliance outcomes, we then computed compliance rates across checklist sections (Section-wise), themes (Theme-wise), and individual criteria (Criterion-wise).

\vspace{2pt}
\subsubsection{\textbf{Findings}}
Table~\ref{tab:compliance-progression} shows how adherence progressed as judgments were consolidated via multi-model consensus, GPT~5 adjudication, and targeted manual review.

\begin{table}[t]
    \centering 
    \caption{Workflow adherence across validation stages}
    \label{tab:compliance-progression}
    \vspace{-8pt}
    \begin{tabular}{p{5.2cm}rc}
    \toprule
    \textbf{Validation Stage} & \textbf{Compliant~} & \textbf{Rate} \\
    \midrule
    Initial 3/4 LLM consensus & 789 / 2,850 & 28\%  \\
    After GPT~5 adjudication & 1,062 / 2,850 & 37\% \\
    After manual review & 1,079 / 2,850 & 38\%  \\
    \bottomrule
    \end{tabular}
    \vspace{-7pt}
\end{table}

\smallskip
\noindent\textbf{Open-source project workflows generally show low adherence to the GHA compliance checklist.} Across 95 workflows and 30 checklist items, only 28\% of items initially achieved consensus among four LLMs. GPT~5 adjudication raised this to 37\%, with manual review contributing a marginal one percentage point (38\%). This indicates that most recommended practices are not followed, revealing widespread CI quality gaps in open-source projects.

\smallskip
\noindent\textbf{Adherence varies across workflow components.} Normalized compliance rates were highest for \texttt{WORKFLOW} and \texttt{JOBS} (40\%) and lowest for \texttt{PERMISSIONS} (4\%), while \texttt{STEPS} achieved 38\%. Although \texttt{STEPS} contained half of all criteria, its normalized adherence closely matched the overall average, suggesting that no single component consistently outperforms others in practice.

\noindent\textbf{Thematic compliance is uneven.} Among themes, \texttt{Clarity} had the highest adherence (68\%), indicating generally acceptable naming, structure, and readability. Adherence was moderate for \texttt{Input Validation} (47\%) and \texttt{Environment} (43\%), while \texttt{Error/Failure Handling} (39\%), \texttt{Modularity} (38\%), and \texttt{Performance} (38\%) were inconsistently implemented. \texttt{Maintainability} (29\%) and \texttt{Security} (26\%) showed the lowest adherence, revealing ongoing weaknesses in robustness, maintainability, and secure configuration.

\smallskip
\noindent\textbf{Even widely recommended practices are rarely applied.} At the criterion level, the most frequently satisfied items were handling workflow failures with notifications (9\%), clear job names (8\%), decomposing complex commands (8\%), validating boot JDK inputs (8\%), and avoiding weak file-change detection (7\%) (Table~\ref{tab:compliance}). This indicates that no checklist item achieves near-universal adoption, even for fundamental practices.

\begin{table*}[ht]
    \centering
    \caption{Most frequently satisfied workflow checklist items. Rate = compliant frequency/total assessed per criterion}
    \vspace{-8pt}
    \label{tab:compliance}
    \resizebox{0.75\textwidth}{!}{
    \begin{tabular}{p{3.5cm} r r r}
    \toprule
    \textbf{Category} & \textbf{Subcategory / Criterion} & \textbf{Compliant Frequency} & \textbf{Rate} \\
    \midrule
    \multirow{4}{*}{Section-wise} 
        & Jobs         & 422  & 40\% \\
        & Workflow     & 114  & 40\% \\
        & Steps        & 539  & 38\% \\
        & Permissions  & 4    & 4\%  \\
    \midrule
    \multirow{8}{*}{Theme-wise}
        & Clarity                  & 129 & 68\%  \\
        & Input Validation         & 178 & 47\%  \\
        & Environment              & 82  & 43\%  \\
        & Error / Failure Handling & 147 & 39\%  \\
        & Modularity               & 143 & 38\%  \\
        & Performance              & 143 & 38\%  \\
        & Maintainability          & 110 & 29\%  \\
        & Security                 & 147 & 26\%  \\
    \midrule
    \multirow{5}{*}{Top 5 Criterion-wise}
        & Workflow handles failures / provides notifications  & 95 & 9\% \\
        & Job names clear and unique                          & 88 & 8\% \\
        & Complex commands split into smaller steps           & 86 & 8\% \\
        & Boot JDK platform inputs validated                  & 85 & 8\% \\
        & Avoid weak file-change detection (e.g., git status) & 74 & 7\% \\
    \bottomrule
    \end{tabular}
    }
    \vspace{-5pt}
\end{table*}

\smallskip
\noindent\textbf{Automated auditing reveals systematic CI gaps.} GPT~5 adjudication improves LLM-based multi-model evaluation, enabling scalable workflow quality assessment with minimal human effort. Applying this approach to open-source workflows shows that, while pipelines are generally readable and structured, they frequently lack robust error handling, modular decomposition, performance optimization, and security controls, highlighting persistent gaps in CI quality and the need for better tooling and community guidance.

\vspace{-10pt}
\begin{rqbox}
\textbf{RQ4 Summary:}
    Open-source GitHub workflows show low adherence to recommended practices, with overall compliance at 28–38\%. Clarity is relatively high (68\%), but security (26\%) and maintainability (29\%) remain weak. Under 10\% follow even the most common practices, indicating persistent gaps in CI robustness, modularity, and security.
\end{rqbox}

\vspace{-5pt}
\section{Discussion}
\label{sec:Discussion}

\noindent\textbf{Compliance is driven by fine-grained, cross-layer decisions.}
RQ1 shows that most compliance criteria target step-level configurations, yet many of these criteria cut across workflow, job, and step boundaries. Security, environment setup, and modularity cannot be assessed in isolation because their correctness depends on how decisions propagate across layers. This explains why syntax-level or section-local checks are insufficient: many violations emerge only when execution logic, runner context, and global settings are considered together. Effective compliance analysis therefore requires cross-section reasoning that captures execution semantics rather than surface structure alone.

\smallskip
\noindent\textbf{LLMs exhibit systematic, not random, inconsistency.}
RQ2 and RQ3 jointly show that disagreement among LLMs is widespread but patterned. Models agree on deterministic, structural criteria such as modularity and naming, but diverge on security, environment, and conditional logic. These disagreements stem from differing implicit assumptions about applicability, risk tolerance, and defaults, not from noise. As a result, LLM judgments are neither interchangeable nor safely composable without explicit resolution mechanisms. Treating LLM outputs as authoritative labels is therefore unjustified for CI compliance tasks.

\smallskip
\noindent\textbf{Adjudication mitigates ambiguity but exposes hard limits.}
RQ3 demonstrates that adjudication with a stronger model resolves a majority of conflicts and substantially increases label stability. However, nearly half of disputed cases still require human judgment, especially where compliance depends on workflow intent, tool semantics, or context-specific security expectations. These cases reveal a fundamental limit of context-free reasoning over YAML: some compliance questions are underspecified without project-level knowledge. Adjudication improves efficiency, but it does not eliminate the need for expert review.

\smallskip
\noindent\textbf{Low compliance reflects Gaps in practice, Not just detection.}
RQ4 shows that adherence to recommended GHA practices is uniformly low across open-source workflows, including for basic security, permissions, and error-handling criteria. Even after adjudication and manual review, no checklist item approaches widespread adoption. This suggests that non-compliance is not primarily an artifact of model disagreement but reflects genuine gaps in CI practice. Workflows tend to be readable and functional, yet systematically under-instrumented for robustness, performance, and security.

\vspace{-3pt}
\subsection{Implications}

\noindent\textbf{For CI Workflow Developers.}
The results show that most compliance violations stem from step-level logic and conditionals, not high-level structure. Developers should prioritize validating inputs, constraining permissions, and making execution intent explicit over simply making workflows runnable. Security and error handling must be encoded in each workflow step, not assumed or inherited. Checklists based on platform documentation provide a practical baseline for self-auditing before using automated tools.

\smallskip
\noindent\textbf{For AI Tool Developers.}
Our results show that LLMs struggle most with context-dependent and non-applicable compliance criteria, causing systematic over-flagging. Auditing tools should expose uncertainty and distinguish mandatory from conditional rules instead of forcing binary decisions, potentially leveraging AI-assisted mechanisms such as CI/CD agents~\cite{ghaleb2026agentscicd} to flag uncertain cases and defer ambiguous decisions. Multi-model adjudication is more reliable than single-model, but must surface unresolved cases for human review rather than hiding disagreement.

\smallskip
\noindent\textbf{For Researchers.}
This study derives 30 compliance criteria from workflow and job documentation, yet many violations occur at the step level and in execution logic. This suggests that CI smell research should therefore look beyond workflow or job structure, even when based on documentation. Future work can refine documentation-driven approaches with finer-grained criteria for command composition, conditional execution, input handling, and security-sensitive steps. These extensions would complement existing smell taxonomies and better reflect where compliance failures actually occur.

\smallskip
\noindent\textbf{For CI Service Providers (e.g., GitHub Actions).}
The low adherence rates indicate that best practices are either hard to discover, hard to apply, or poorly enforced. CI providers can improve compliance by offering stronger defaults, clearer normative guidance, and first-class validation for security and permissions. Native tooling that explains why a configuration is risky or non-compliant, rather than merely flagging it, would reduce ambiguity for both humans and automated auditors.

\section{Threats to Validity}
\label{sec:Threats to Validity}

\smallskip
\noindent\textbf{Construct Validity.}
The compliance checklist was derived from official GitHub Actions documentation and refined through iterative application to real workflows. This grounding reduces the risk of inventing ad hoc or undocumented criteria. However, the checklist focuses primarily on workflow- and job-related guidance, even though many violations manifest at the step level. While this was an intentional design choice to anchor criteria in documented platform guarantees, it may underrepresent undocumented or community-driven practices.
Some criteria are conditionally applicable (e.g., caching, security scanning, input validation), and applicability depends on workflow intent that is not always explicit in YAML. Although the \textit{N/A} option was used to mitigate this issue, misclassification of applicability remains possible and may contribute to disagreement. Finally, some YAML files represent partial CI configurations (e.g., reusable workflows or auxiliary pipelines), which may appear non-compliant when assessed in isolation despite being correct in the project context.

\smallskip
\noindent\textbf{Internal Validity.}
LLM outputs may be influenced by prompt formulation, model defaults, or undocumented heuristics. We mitigated this by using a fixed zero-shot prompt, deterministic decoding for open-weight models, and structured JSON outputs. Still, different prompting strategies or few-shot examples 
might yield different results, and prompt sensitivity remains an avenue for future investigation.
Manual adjudication introduces potential reviewer bias. To reduce this threat, we used stratified sampling with full criterion coverage, blinded review, and documentation-backed resolution of disagreements. GPT~5 was used only as a dispute resolver and not treated as ground truth. Nevertheless, some borderline cases remain inherently subjective, especially for security and performance practices, which could affect adjudication outcomes.

\smallskip
\noindent\textbf{External Validity.}
The study focuses on Java-based open-source projects using GitHub Actions. While this controls for ecosystem variability, results may differ for other languages, build tools, or private repositories with stricter security policies. The selected LLMs represent diverse open-weight architectures plus a stronger proprietary model, but findings may not generalize to other models or future versions.
The workflows analyzed are from public repositories with minimum activity thresholds, likely biasing the sample toward better-maintained projects. This means the low compliance rates observed are probably conservative estimates rather than overstatements.

\smallskip
\noindent\textbf{Conclusion Validity.}
Agreement metrics and compliance rates depend on how \textit{N/A} cases are handled and how split decisions are resolved. While we report intermediate agreement distributions and use multiple resolution paths (consensus, adjudication, manual review), alternative aggregation strategies could yield slightly different absolute rates. However, the main conclusions—systematic LLM disagreement, concentration of violations at fine-grained levels, and low overall adherence—are robust across resolution stages.

\vspace{4.5pt}
\section{Conclusion}
\label{sec:Conclusion}
This paper introduced a documentation-grounded checklist of 30 compliance criteria for auditing GitHub Actions (GHA) workflows and evaluated the consistency of large language models (LLMs) in applying it. The criteria were derived from official GHA documentation and focused on workflow compliance, providing a reproducible basis for automated auditing.
Our evaluation shows substantial variation across models. \textsc{Gemma-3~12B} achieved the highest agreement rate (90\%), while \textsc{LLaMA-3.1~8B} performed lowest (61\%). Overall agreement was limited, with Fleiss’ $\kappa$ = 0.28, indicating only \emph{fair} consistency. Split cases (35\%) revealed recurring challenges in workflow reasoning, including job dependencies, secrets usage, and caching behavior.
Using GPT~5 as an adjudicator resolved 57\% of split cases and reduced manual effort, but human review remained necessary for ambiguous or context-dependent decisions. Overall, the results suggest that LLMs can support CI compliance auditing, but reliable use requires a hybrid setup combining automated reasoning with expert adjudication.
For CI developers, the checklist can support the design and review of GitHub Actions workflows. For AI tool builders, the results highlight the need for stronger documentation grounding and better reasoning over cross-job dependencies. For researchers, the findings suggest defining compliance criteria directly from documentation and extending analysis beyond isolated jobs to workflow-level properties. For CI providers such as GitHub Actions, the criteria could support native linting and auditing features to improve clarity and trust in automated checks.

\medskip
\noindent\textbf{Future work.}
Future work should extend the checklist to incorporate additional criteria about workflow structure, triggers, built-in features, and reusable components. We plan to evaluate our work on a larger and more diverse set of workflows across programming languages, repository types, and CI frameworks. We also aim to improve LLM grounding through fine-tuning and retrieval over official CI documentation to reduce disagreement and hallucination. Finally, we will conduct cross-project analyses to study variations in compliance patterns and LLM performance across domains and development practices.

\vspace{4.5pt}
\section*{Artifact Availability}
A replication package (scripts, data, and raw results) used to produce the findings of this study is available online on GitHub~\cite{our_replication_package}.

\vspace{4.5pt}
\begin{acks}
This work is funded by the Natural Sciences and Engineering Research Council of Canada (NSERC): RGPIN-2025-05897.
\end{acks}

\clearpage
\balance
\bibliographystyle{ACM-Reference-Format}
\bibliography{paper}

@article{stureborg2024large,
  title={Large language models are inconsistent and biased evaluators},
  author={Stureborg, Rickard and Alikaniotis, Dimitris and Suhara, Yoshi},
  journal={arXiv preprint arXiv:2405.01724},
  year={2024}
}

@article{guo2025repoaudit,
  title={Repoaudit: An autonomous llm-agent for repository-level code auditing},
  author={Guo, Jinyao and Wang, Chengpeng and Xu, Xiangzhe and Su, Zian and Zhang, Xiangyu},
  journal={arXiv preprint arXiv:2501.18160},
  year={2025}
}

@misc{openai_gpt5_2025,
    author = {{OpenAI}},
    title = {Introducing GPT-5},
    year = {2025},
    publisher = {OpenAI},
    url = {https://openai.com/gpt-5},
    urldate = {2025-08-27},
    note = {Online; accessed 27-August-2025}
}

@inproceedings{beller2017travistorrent,
  title={Travistorrent: Synthesizing {Travis CI} and {GitHub} for full-stack research on continuous integration},
  author={Beller, Moritz and Gousios, Georgios and Zaidman, Andy},
  booktitle={2017 IEEE/ACM 14th International Conference on Mining Software Repositories},
  pages={447--450},
  year={2017},
  organization={IEEE}
}

@inproceedings{chopra2025multici,
  title={From First Use to Final Commit: Studying the Evolution of Multi-CI Service Adoption},
  author={Chopra, Nitika and Ghaleb, Taher A},
  booktitle={International Conference on Software Maintenance and Evolution},
  pages={773--778},
  year={2025},
  organization={IEEE}
}

@Misc{our_replication_package,
  author={Edward Abrokwah and Taher A. Ghaleb},
  title = {Auditing {GitHub Actions} Workflows: A Compliance Checklist and Evaluation Using {LLMs}~({Replication Package})},
  year={2026},
  howpublished = {\url{https://github.com/Taher-Ghaleb/GHACompliance-EASE2026}}
}

@article{zampetti2020empirical,
  title={An empirical characterization of bad practices in continuous integration},
  author={Zampetti, Fiorella and Vassallo, Carmine and Panichella, Sebastiano and Canfora, Gerardo and Gall, Harald and Di Penta, Massimiliano},
  journal={Empirical Software Engineering},
  volume={25},
  number={2},
  pages={1095--1135},
  year={2020},
  publisher={Springer}
}

@article{gallaba2018use,
  title={Use and misuse of continuous integration features: An empirical study of projects that (mis) use Travis CI},
  author={Gallaba, Keheliya and McIntosh, Shane},
  journal={IEEE Transactions on Software Engineering},
  volume={46},
  number={1},
  pages={33--50},
  year={2018},
  publisher={IEEE}
}

@article{santana2025evaluating,
  title={Evaluating LLMs Effectiveness in Detecting and Correcting Test Smells: An Empirical Study},
  author={Santana Jr, Enio G and Junior, Jander Pereira Santos and Almeida, Erlon P and Ahmed, Iftekhar and Neto, Paulo Anselmo da Mota Silveira and de Almeida, Eduardo Santana},
  journal={arXiv preprint arXiv:2506.07594},
  year={2025}
}

@article{taibi2017developers,
  title={How developers perceive smells in source code: A replicated study},
  author={Taibi, Davide and Janes, Andrea and Lenarduzzi, Valentina},
  journal={Information and Software Technology},
  volume={92},
  pages={223--235},
  year={2017},
  publisher={Elsevier}
}

@article{abdin2024phi,
  title={Phi-4 technical report},
  author={Abdin, Marah and Aneja, Jyoti and Behl, Harkirat and Bubeck, S{\'e}bastien and Eldan, Ronen and Gunasekar, Suriya and Harrison, Michael and Hewett, Russell J and Javaheripi, Mojan and Kauffmann, Piero and others},
  journal={arXiv preprint arXiv:2412.08905},
  year={2024}
}

@article{team2025gemma,
  title={Gemma 3 technical report},
  author={Team, Gemma and Kamath, Aishwarya and Ferret, Johan and Pathak, Shreya and Vieillard, Nino and Merhej, Ramona and Perrin, Sarah and Matejovicova, Tatiana and Ram{\'e}, Alexandre and Rivi{\`e}re, Morgane and others},
  journal={arXiv preprint arXiv:2503.19786},
  year={2025}
}

@online{mistralai2024mistral7b,
  author = {{Mistral AI}},
  title = {Mistral 7B},
  year = {2024},
  url = {https://mistral.ai/news/mistral-7b/},
  note = {Accessed: 2025-10-18}
}

@article{touvron2023llama,
  title={Llama: Open and efficient foundation language models},
  author={Touvron, Hugo and Lavril, Thibaut and Izacard, Gautier and Martinet, Xavier and Lachaux, Marie-Anne and Lacroix, Timoth{\'e}e and Rozi{\`e}re, Baptiste and Goyal, Naman and Hambro, Eric and Azhar, Faisal and others},
  journal={arXiv preprint arXiv:2302.13971},
  year={2023}
}

@incollection{marcondes2025using,
  title={Using {Ollama}},
  author={Marcondes, Francisco S and Gala, Adelino and Magalh{\~a}es, Renata and Perez de Britto, Fernando and Dur{\~a}es, Dalila and Novais, Paulo},
  booktitle={Natural Language Analytics with Generative Large-Language Models: A Practical Approach with Ollama and Open-Source LLMs},
  pages={23--35},
  year={2025},
  publisher={Springer}
}

@inproceedings{hilton2017trade,
  title={Trade-offs in continuous integration: assurance, security, and flexibility},
  author={Hilton, Michael and Nelson, Nicholas and Tunnell, Timothy and Marinov, Darko and Dig, Danny},
  booktitle={Proceedings of the 11th Joint Meeting on Foundations of Software Engineering},
  pages={197--207},
  year={2017}
}

@misc{Fowler_CI, 
  title={Continuous Integration},
  author={Martin Fowler}, 
  howpublished = "\url{https://martinfowler.com/articles/originalContinuousIntegration.html}"
}

@inproceedings{wang2024large,
  title={Large language models are not fair evaluators},
  author={Wang, Peiyi and Li, Lei and Chen, Liang and Cai, Zefan and Zhu, Dawei and Lin, Binghuai and Cao, Yunbo and Kong, Lingpeng and Liu, Qi and Liu, Tianyu and others},
  booktitle={Proceedings of the 62nd Annual Meeting of the Association for Computational Linguistics},
  pages={9440--9450},
  year={2024}
}

@article{zheng2023judging,
  title={Judging {LLM-as-a-Judge} with {MT-Bench} and chatbot arena},
  author={Zheng, Lianmin and Chiang, Wei-Lin and Sheng, Ying and Zhuang, Siyuan and Wu, Zhanghao and Zhuang, Yonghao and Lin, Zi and Li, Zhuohan and Li, Dacheng and Xing, Eric and others},
  journal={Advances in neural information processing systems},
  volume={36},
  pages={46595--46623},
  year={2023}
}

@inproceedings{liu2024llms,
  title={{LLMs} as narcissistic evaluators: When ego inflates evaluation scores},
  author={Liu, Yiqi and Moosavi, Nafise Sadat and Lin, Chenghua},
  booktitle={Findings of the Association for Computational Linguistics: ACL 2024},
  pages={12688--12701},
  year={2024}
}

@inproceedings{valenzuela2024hidden,
  title={The Hidden Costs of Automation: An Empirical Study on {GitHub Actions} Workflow Maintenance},
  author={Valenzuela-Toledo, Pablo and Bergel, Alexandre and Kehrer, Timo and Nierstrasz, Oscar},
  booktitle={2024 IEEE International Conference on Source Code Analysis and Manipulation (SCAM)},
  pages={213--223},
  year={2024},
  organization={IEEE}
}

@misc{static_analysis_limits,
  title = {Static Code Analysis: Top 7 Methods, Pros/Cons and Best Practices},
  author = {{Oligo Security}},
  year = {2025},
  howpublished = {\url{https://www.oligo.security/academy/static-code-analysis}},
  note={Accessed on Jan 20, 2026}
}

@misc{codecov_breach,
  title = {Codecov Bash Uploader Security Incident},
  author = {{Codecov}},
  year = {2021},
  howpublished = {\url{https://about.codecov.io/security-update}},
  note = {Accessed: 2026-01-20}
}

@misc{cycode_gha_vulns,
  title = {How we found vulnerabilities in {GitHub Actions} {CI/CD} pipelines},
  author = {{Cycode Security Research}},
  year = {2024},
  howpublished = {\url{https://cycode.com/blog/github-actions-vulnerabilities}},
  note = {Accessed: 2026-01-20}
}

@misc{legit_workflow_run,
  title = {Vulnerable {GitHub Actions} Workflows Part 1: Privilege Escalation Inside Your {CI/CD} Pipeline},
  author = {{Legit Security Research Team}},
  year = {2024},
  howpublished = {\url{https://www.legitsecurity.com/blog/github-privilege-escalation-vulnerability}},
  note = {Accessed: 2026-01-20}
}

@misc{linter_semantic_gap,
  title = {What Is the Difference Between Static Code Analysis and Linting?},
  author = {{IN-COM DATA SYSTEMS}},
  year = {2025},
  howpublished = {\url{https://www.in-com.com/blog/what-is-the-difference-between-static-code-analysis-and-linting}},
  note = {Accessed: 2026-01-20}
}

@inproceedings{amiri2025enhancing,
  title={Enhancing answer reliability through inter-model consensus of large language models},
  author={Amiri-Margavi, Alireza and Jebellat, Iman and Jebellat, Ehsan and Davoudi, Seyed Pouyan Mousavi},
  booktitle={IFIP International Conference on Artificial Intelligence Applications and Innovations},
  pages={299--316},
  year={2025},
  organization={Springer}
}

@inproceedings{wagner2025towards,
  title={Towards evaluation guidelines for empirical studies involving {LLMs}},
  author={Wagner, Stefan and Bar{\'o}n, Marvin Mu{\~n}oz and Falessi, Davide and Baltes, Sebastian},
  booktitle={2025 IEEE/ACM International Workshop on Methodological Issues with Empirical Studies in Software Engineering (WSESE)},
  pages={24--27},
  year={2025},
  organization={IEEE}
}

@misc{gha_docs,
  title = {{GitHub Actions} Documentation},
  author = {{GitHub}},
  howpublished = {\url{https://docs.github.com/en/actions}},
  note = {Accessed: 2025-11-17}
}

@inproceedings{stol2016grounded,
  title={Grounded theory in software engineering research: a critical review and guidelines},
  author={Stol, Klaas-Jan and Ralph, Paul and Fitzgerald, Brian},
  booktitle={Proceedings of the 38th International conference on software engineering},
  pages={120--131},
  year={2016}
}

@article{gupta2026reliabilitybench,
  title={{ReliabilityBench}: Evaluating {LLM} Agent Reliability Under Production-Like Stress Conditions},
  author={Gupta, Aayush},
  journal={arXiv preprint arXiv:2601.06112},
  year={2026}
}

@inproceedings{khatami2024catching,
  title={Catching smells in the act: A {GitHub Actions} workflow investigation},
  author={Khatami, Ali and Willekens, C{\'c}dric and Zaidman, Andy},
  booktitle={2024 IEEE International Conference on Source Code Analysis and Manipulation (SCAM)},
  pages={47--58},
  year={2024},
  organization={IEEE}
}

@inproceedings{zhang2022buildsonic,
  title={BuildSonic: Detecting and repairing performance-related configuration smells for continuous integration builds},
  author={Zhang, Chen and Chen, Bihuan and Hu, Junhao and Peng, Xin and Zhao, Wenyun},
  booktitle={Proceedings of the 37th IEEE/ACM international conference on automated software engineering},
  pages={1--13},
  year={2022}
}

@inproceedings{ghaleb2025llm4ci,
  title={Can {LLMs} Write {CI}? A Study on Automatic Generation of GitHub Actions Configurations},
  author={Ghaleb, Taher A and Rathnayake, Dulina},
  booktitle={2025 IEEE International Conference on Software Maintenance and Evolution},
  pages={767--772},
  year={2025},
  organization={IEEE}
}

@article{hossain2025cigrate,
  title={{CIgrate}: Automating {CI} Service Migration with Large Language Models},
  author={Hossain, Md Nazmul and Ghaleb, Taher A},
  journal={arXiv preprint arXiv:2507.20402},
  year={2025}
}

@inproceedings{xu2025logsage,
  title={LogSage: An {LLM}-Based Framework for {CI/CD} Failure Detection and Remediation with Industrial Validation},
  author={Weiyuan Xu and Juntao Luo and Tao Huang and Kaixin Sui and Jie Geng and Qijun Ma and Isami Akasaka and Xiaoxue Shi and Jing Tang and Peng Cai},
  booktitle={40th IEEE/ACM International Conference on Automated Software Engineering},
  year={2025}
}

@article{mehta2023automated,
  title={Automated devops pipeline generation for code repositories using large language models},
  author={Mehta, Deep and Rawool, Kartik and Gujar, Subodh and Xu, Bowen},
  journal={arXiv preprint arXiv:2312.13225},
  year={2023}
}

@inproceedings{chomkatek2025decoding,
  title={Decoding CI/CD Practices in Open-Source Projects with LLM Insights},
  author={Chom{\k{a}}tek, {\L}ukasz and Papuga, Jakub and Nowak, Przemyslaw and Poniszewska-Mara{\'n}da, Aneta},
  booktitle={Proceedings of the 33rd ACM International Conference on the Foundations of Software Engineering},
  pages={1638--1644},
  year={2025}
}

@misc{openai_reasoning,
  author       = {{OpenAI}},
  title        = {{Reasoning Models Guide}},
  howpublished = {\url{https://developers.openai.com/api/docs/guides/reasoning}},
  year         = {2026},
  note         = {Accessed:2026}
}

@article{pipis2025wait,
  title={Wait, Wait, Wait... Why Do Reasoning Models Loop?},
  author={Pipis, Charilaos and Garg, Shivam and Kontonis, Vasilis and Shrivastava, Vaishnavi and Krishnamurthy, Akshay and Papailiopoulos, Dimitris},
  journal={arXiv preprint arXiv:2512.12895},
  year={2025}
}

@article{blackwell2024towards,
  title={Towards reproducible llm evaluation: Quantifying uncertainty in llm benchmark scores},
  author={Blackwell, Robert E and Barry, Jon and Cohn, Anthony G},
  journal={arXiv preprint arXiv:2410.03492},
  year={2024}
}

@article{ghaleb2025android,
  title={{CI/CD} Configuration Practices in Open Source {Android} Apps: An Empirical Study},
  author={Ghaleb, Taher A and Abduljalil, Osamah and Hassan, Safwat},
  journal={ACM Transactions on Software Engineering and Methodology},
  volume={35},
  number={2},
  pages={1--40},
  year={2026},
  publisher={ACM New York, NY}
}

@article{ghaleb2019duration,
  title={An empirical study of the long duration of continuous integration builds},
  author={Ghaleb, Taher Ahmed and da Costa, Daniel Alencar and Zou, Ying},
  journal={Empirical Software Engineering},
  volume={24},
  number={4},
  pages={2102--2139},
  year={2019},
  publisher={Springer}
}

@article{ghaleb2019noise,
  title={Studying the Impact of Noises in Build Breakage Data},
  author={Ghaleb, Taher Ahmed and da Costa, Daniel Alencar and Zou, Ying and Hassan, Ahmed E},
  journal={IEEE Transactions on Software Engineering},
  pages={1--14},
  year={2019},
  DOI={10.1109/TSE.2019.2941880},
  publisher={IEEE}
}

@article{ghaleb2022interplay,
  title={Studying the interplay between the durations and breakages of continuous integration builds},
  author={Ghaleb, Taher A and Hassan, Safwat and Zou, Ying},
  journal={IEEE Transactions on Software Engineering},
  volume={49},
  number={4},
  pages={2476--2497},
  year={2022},
  publisher={IEEE}
}

@inproceedings{ghaleb2026agentscicd,
  title={When {AI} Agents Touch {CI/CD} Configurations: Frequency and Success},
  author={Ghaleb, Taher A},
  booktitle={Proceedings of the 23rd International Conference on Mining Software Repositories},
  year={2026},
  pages={1--5},
  organization={ACM}
}

@inproceedings{zhou2026roleci,
  title={Role of {CI} Adoption in Mobile App Success: An Empirical Study of Open-Source {Android} Projects},
  author={Zhou, Xiaoxin and Ghaleb, Taher A and Hassan, Safwat},
  booktitle={Proceedings of the 23rd International Conference on Mining Software Repositories},
  year={2026},
  pages={1--12},
  organization={ACM}
}

@inproceedings{vassallo2020configuration,
  title={Configuration smells in continuous delivery pipelines: a linter and a six-month study on GitLab},
  author={Vassallo, Carmine and Proksch, Sebastian and Jancso, Anna and Gall, Harald C and Di Penta, Massimiliano},
  booktitle={Proceedings of the 28th ACM Joint Meeting on European Software Engineering Conference and Symposium on the Foundations of Software Engineering},
  pages={327--337},
  year={2020}
}

\end{document}